
%

\magnification=1200
\baselineskip=20pt
\normallineskip=8pt
\overfullrule=0pt
\vsize=23 true cm
\hsize=15 true cm
\font\bigfont=cmr10 scaled\magstep1
\font\ninerm=cmr9
\footline={\hss\tenrm\folio\hss}
\pageno=1
\newcount\fignumber
\fignumber=0
\def\fig#1#2{\advance\fignumber by1
 \midinsert \vskip#1truecm \hsize14truecm
 \baselineskip=15pt \noindent
 {\ninerm {\bf Figure \the\fignumber} #2}\endinsert}

\def\ref#1{$^{[#1]}$}
\def\sqr#1#2{{\vcenter{\vbox{\hrule height.#2pt
   \hbox{\vrule width.#2pt height#1pt \kern#1pt
   \vrule width.#2pt}\hrule height.#2pt}}}}

\bigskip
\bigskip
\centerline{\bigfont A BEAM MODEL FOR HYDRAULIC FRACTURING}
\bigskip
\bigskip
\bigskip
\bigskip
\smallskip
\centerline{{\bf F. Tzschichholz$^{1,2,3}$, H.J. Herrmann$^{2}$,
 H. E. Roman$^{3,4}$ and M. Pfuff$^{1}$} }
\medskip
\centerline{$^1$ Institut f\"ur Werkstofforschung}
\centerline{GKSS-Forschungszentrum, Postfach 1160, D-21502 Geesthacht, Germany}
\smallskip
\centerline{$^{2}$ HLRZ, KFA J\" ulich, Postfach 1913,
D-52428 J\" ulich, Germany}
\smallskip
\centerline{$^{3}$ I. Institut f\"ur Theoretische Physik}
\centerline{ Universit\"at Hamburg, D-20355 Hamburg, Germany}
\smallskip
\centerline{$^{4}$Dipartimento di Fisica, Universit\`a di Milano}
\centerline{Via Celoria 16, I-20133 Milano, Italy}
\smallskip
\bigskip
\bigskip
\bigskip
\bigskip
\bigskip
\bigskip
\noindent{\bf Abstract} \par
\smallskip
We investigate numerically the shape of cracks obtained
in hydraulic fracturing at constant pressure using
a square lattice beam model with disorder.
We consider the case in which only beams under tension can break, and
discuss the conditions under which
the resulting cracks may develop fractal patterns.
We also determine the opening volume of the crack and the
elastic stress field in the bulk, quantities which are accessible
experimentally.
\bigskip
\bigskip
\leftline{\bf PACS numbers: } 46.30, 91.60.-x

\vfill\eject
Hydraulic fracturing is a technique of great importance
in soil mechanics and is used systematically to improve
oil recovery. In particular, it has a considerable importance
for the functioning of geothermal wells\ref {1}.
An incompressible fluid, in general water, is pushed with
pressure deep inside a solid, in the case of soil by
injecting it into a deep perforation. The fluid penetrates
into the solid by opening long cracks that radially
go from the injection hole into the material.

In two dimensions controlled experiments have been
performed recently\ref {2}. Water
or air is pushed from above into the center of a
circular Hele Shaw cell filled with clay. While on long
time scales clay behaves like a fluid for high
injection pressures it fractures like a solid (viscoelastic
fluid). In this fracturing regime the resulting cracks
display a disordered ramified structure which appears to obey
self-similarity with a fractal dimension of 1.4 - 1.5.
Compared to the fractal structures
observed in Laplacian systems (dielectric breakdown,
viscous fingering or diffusion-limited aggregation (DLA) )
not only the fractal
dimension is lower but also the angles between branches
are about three times larger (i.e. about $90^{\circ }$)
and no tip splitting is observed.

These interesting observations remain largely unexplained.
While for the Laplacian case the underlying instabilities
(Saffman-Taylor, tip splitting, side branching) have been
heavily investigated\ref {3} and numerical clusters of tens of millions
of particles have been generated\ref {4}, in the case of fracture
much work still needs to be done. A recent stability analysis
of the shape of a circular hole with either internal pressure
or in a stretched membrane has shown that contrary to the common belief
a large difference can be expected in the shape of a crack
between the two cases\ref {5}. The origin of this
is the non-linear dependence of the growth velocity of the crack
surface due to the threshold in cohesion force that must be overcome
to break the material. In the Laplacian case these differences
do not exist in the limit of large clusters\ref {6}.

Numerically the breaking of a material from a central hole
was first investigated by Louis and Guinea\ref {7} using a
triangular network of springs and stretching the network
radially on the outer boundary into the six directions of a
hexagon. They observed fractal cracks having a fractal
dimension close to that of DLA. These findings were studied
subsequently by various groups\ref {8}. It became clear
that the patterns depend very much on the type of applied
displacements (shear, uni-axial, radial). Since networks of
springs need more extended structures to transfer momenta
it turned out to be more efficient to use a beam model\ref {9}
which is no longer a central force model.
For more details on the work that has been done we refer to
the book of Ref.~10.

While in previous works imposed displacements were applied only
at the external surface of the solid
(Dirichlets boundary value problem),
in this paper we introduce and study a model in which
the imposed load represents a pressure that acts
along the whole (inner) surface of the crack in a direction
perpendicular to the surface (von Neumann
boundary value problem).
In this way, the
point of application of the imposed load varies during the
growth of the crack, a situation that more realistically
describes the case of hydraulic fracturing.


We consider the beam model (as defined in p.~232 of Ref.~10) on
a square lattice of linear size $L$. Each of the
lattice sites $i$ carries three real variables:
the two translational displacements $x_i$ and $y_i$ and
a rotational displacement $\varphi_i $. Next neighbouring sites
are rigidly connected by elastic beams of length $l$\ref {11}.
When a site is rotated $\varphi_i \ne 0$
then the beams must bend accordingly (Fig. 1a) and continue to
tangentially form $90^{\circ}$ angles with each other.
In this way local momenta are taken into account.

Since most materials typically crack under tension and
in much less degree under compression we will assume that
only beams that are under tension are allowed to break,
i.e. to be irreversibly removed from the
forces and momenta balance equations\ref {12}.
This means that
the cohesion force against
compression is actually infinite, i.e. the beams can be compressed but
not broken by compression.

At the place into which the incompressible
fluid is supposed to be injected (center of the lattice)
one vertical beam connecting sites $i$ and $j$ is removed.
Since we want to simulate the loading of a crack by the
injected fluid
an invariant double force
${\vec F}_i$ and ${\vec F}_j$
conserving momentum
is applied at the sites $i$ and $j$
pointing from the hole (removed beam) into the elastic bulk (Fig.1b).
Thus, force densities are replaced by discrete force vectors
${\vec F} = (F_x, F_y)$,
$$ {\vec F}_i = F_0 ~(0,1),~~{\vec F}_j = -F_0 ~(0,1).   \eqno(1) $$
The forces for a broken horizontal bond are defined correspondingly.
The acting pressure $P$ is just the force
$F_0$ per beam thickness $d$ and beam length $l$ and its value is kept
fixed during the fracture process.

Each time a beam is removed a new force dipole is applied at
its corresponding nearest-neighbor sites
$i$ and $j$ which destroys the balance of forces existing previously.
Consequently the lattice must be relaxed
and the new static equilibrium must be obtained again. This is done
in our case using conjugate gradient\ref {13}.
After the unique displacement field ${\vec u}_i$ corresponding
to the new boundary conditions is determined
one can decide which is the next beam to be broken.
Due to this procedure we only consider the case of an
instantaneous relaxation, i.e. we assume
that the physical process of stress redistribution is much
faster than the actual fracture process.

For simplicity, only beams along the surface of the inner hole
are considered in the breaking process\ref {14}.
For each of these beams the force $f_{ij}$
acting along its axis is determined
and only if this $f_{ij}$ is positive, i.e. a tension, the
beam can be broken (active beams).
In addition, since we are interested in random structures we need
to incorporate some stochastic mechanism or disorder in the model.
The disorder may be either quenched, i.e. the beams may have different
thresholds for breaking,
or it may be annealed, i.e. one selects the next broken beam randomly
according to some probability. In the following we will employ the
second procedure.

Among the active beams one is chosen with a probability $p_{ij}$,
which is determined in analogy to dielectric breakdown\ref {15} to be
$$p_{ij} \propto f_{ij}^{\eta}  \eqno(2)$$
if $f_{ij}$ is larger than a certain cohesion force
threshold $f_{\rm coh}\ge 0$ and zero otherwise\ref {16}.
Here, $0\leq\eta<\infty$
is a parameter which determines the {\it local} breaking properties of the
material. If $\eta\to\infty$ only the most stretched beam breaks,
i.e. the one with the largest $f_{ij}$. The opposite case corresponds
to $\eta=0$ where all active beams can break with
the same probability, i.e. to a situation in which
heterogeneities are more important than tension forces.


Let us start discussing our results for the case $\eta=1$
(similar results are obtained for $\eta>1$) and $f_{\rm coh}=0$.
For the Laplacian case, the value $\eta=1$ corresponds to DLA
with a fractal dimension $d_f\cong 1.7$ in two dimensions\ref {15}. For
the present beam model, however, topologically linear cracks develop
($d_f=1$) as shown in Fig. 2a. This indicates that
the distribution of $p_{ij}$ is much narrower than for DLA
and essentially only the
most stretched beams (located at the tips of the crack) break.
Thus, already for $\eta=1$,
only a single branch persist on large length scales.
The colors in Fig.2 represent the stress field. They show
that the strongest gradients are indeed around the tips and that regions
other than the tips are under constant compression of magnitude $P$.

Also surprising is the result obtained when $\eta=0$,
i.e. the case in which all beams under
tension can be selected for breaking with the same
probability. For the Laplacian case, one obtains compact clusters
of spherical shape and fluctuations occur only at their surface\ref {15}.
In the present case, the resulting cracks (shown in Fig. 2b)
display ramifications
which persist up to scales comparable to the system size $L$,
and it seems plausible that they might be fractal.
The colors show that the strongest gradients occur again around the
crack tips and that the regions between crack branches
are virtually under constant compression of magnitude $P$.
One clearly sees the opening of the crack being larger
on certain main cracks than on the side branches.
Although only weakly visible the
system of Fig.~2b is no perfect square anymore but has
small bulges\ref {17}.



To characterize quantitatively the shape of the cracks
obtained when $\eta=0$, we have
plotted the number $N$ of broken beams against the radius of
gyration $R_G$ of the crack\ref {18}.
The results are shown in the log-log plot of Fig.~3 for the case
of free external boundaries.
We see a powerlaw behaviour $N\sim R_G^{d_f}$ over nearly
two decades and the slope gives a fractal
dimension of $d_f = 1.56 \pm 0.05$. Similar results were obtained for
the case of periodic boundary conditions.

Our model is different with respect to the Laplacian one
in that all the beams under
compression cannot break. The fact that in our case
the cracks become fractal when $\eta=0$ shows that only at very
few points, namely at the crack tips, the crack surface
is under tension and everywhere else under compression.


It is also interesting to consider a fractal structure
through its intrinsic metric by using as geodesics
only the shortest paths that are entirely on the cluster.
In this topological or ``chemical'' space \ref {19}, one can
calculate the number of broken bonds $N_{\ell}$ that are found within
a chemical distance $\ell$ \ref {20}.
We have obtained a powerlaw behavior
$N_{\ell}\sim \ell^{d_\ell}$ with $d_\ell=1.4\pm 0.05$,
for the same cracks considered in Fig. 3, somewhat
smaller than $d_f$.
The interest in the ``chemical dimension'' $d_{\ell}$ is because
for loopless structures, such as the cracks described here,
both $d_f$ and
$d_{\ell}$ completely determine the scaling behavior of
diffusion controlled transport phenomena \ref {21} such
as e.g. the diffusion of chemical substances which can be present
in the fluid contained within the cracks.

For a finite cohesion threshold $f_{\rm coh}>0$ and $\eta=0$ the cracks
tend to grow more anisotropically since, initially,
only the most stressed beams can break\ref {16}.
However, much larger system sizes
than the ones considered here are required in order to
reach the asymptotic shape of the cracks when $f_{\rm coh}>0$.
This is because a finite ratio $f_{\rm coh}/P$
introduces an additional length scale in the system\ref {22}, and
the scaling behavior obtained when $f_{\rm coh}=0$ is recovered only
asymptotically.

Our simulations are performed at constant pressure and
the volume $V$ of the crack opening which corresponds to the
amount of fluid that has penetrated into the soil is
a measurable variable. We measure this volume by taking into account
the actual displacements of the lattice sites as the crack grows.
These displacements are given from the
elastic solution ${\vec u}_i=(x_i,y_i,\varphi_i)$,
and the volume elements $\Delta V_{ij}$
connecting sites $i$ and $j$ are obtained simply as
$\Delta V_{ij}=(x_i-x_j)l d$ for horizontal broken beams, with $l$
and $d$
being the beam length and thickness, respectively.
Similar expressions hold for vertical broken beams. The total
crack opening volume $V$ is just
the sum of all the volume elements $\Delta V_{ij}$.

In Fig.~4 we show the crack volume $V$
for free and periodic boundary conditions as a function of the
radius of gyration $R_G$ in a log-log plot.
The slope gives an exponent consistent with two, i.e. the
spatial dimension. This agrees with the observation that a
finite amount of fluid actually enters into the hole.
When the crack approaches the boundary
too much the curve has an artificial increase in slope due
to boundary effects.
For comparison, we also measured the volume of a
deterministic straight crack as shown in Fig.~4. For such a crack in an
infinite plate
it is well known from elasticity theory that the crack opening
volume is proportional to the squared crack length
(in two dimensions) \ref {23}.
This shows that (1) although the crack surface
(number of broken beams) is fractal the volume of the crack is
not, (2) the relationship $V\sim R_G^2$ is {\it independent} of the
crack structure itself and (3) particular external boundary
conditions play a minor role as long as the ``typical length''
$R_G$ describing the crack structure is much smaller than the system
size.


\bigskip
We have presented a discrete fracture model in two dimensions
adapted to the conditions of hydraulic cracking of soils
as typically used for instance to create additional heat exchange
surface for geothermal energy recovery.
Only tensile forces break the solid and the heterogeneities
are considered to be dominant \ref {24}.
Under these circumstances the
crack patterns are fractal and we determined the fractal dimension
to be less than that of DLA.
Our results for $\eta = 0$ reproduce some of the conspicuous features
observed in the 2d experiments of Orl\'eans for high injection rates
and rigid pastes\ref {2}:
The crack patterns are more kinky than DLA
and the fractal dimension is lower, their (large) error
actually overlap.
Thus, the basic
assumptions of our model, namely that the material does
not open under compression and that for $\eta=0$ the
heterogeneities are more relevant than the actual value of the tension
force, may describe the case of pastes
with large clay/water ratios used in Orl\'eans.

We also have clear numerical evidence that at constant pressure
the opening volume of the crack grows proportionaly to the
area spanned by the crack, i.e. the square of its radius
of gyration, {\it independent} of the crack structure itself.

Our model should be considered as yet quite preliminary
concerning the understanding of real industrial hydrostatic
fracturing. On one hand real soils are three dimensional
and in fact we do not know whether in three dimensions the patterns
are finger-like as in two dimensions or rather consisting
of randomly attached fracture planes. The later case seems
more likely has yet never been seen numerically up to now.
Also the role of the heterogeneities must be investigated closer.
In soils the disorder is quenched, i.e. the randomness (in breaking
threshold, modulus, etc) are assigned before cracking starts.
We have used ``annealed'' randomness: random numbers are used
during the breaking process in order to select beams with a certain
probability. Also realistic, e.g. Weibull distributions,
should be considered.
Other important ingredients that should be taken into account
when trying to make the model more realistic are plasticity,
pressure gradients and hydrodynamic effects in the fluid,
stress corrosion and short time effects, like shock waves.

\bigskip
\bigskip
\bigskip
\bigskip
\leftline{\bf Acknowledgements}
\smallskip
H.E.R. gratefully acknowledges financial support from the
Alexander von Humboldt Stiftung (Feodor Lynen program).

\bigskip
\noindent{\bf References}
\medskip
\item{1.} H. Takahashi and H. Ab\'e, in {\it Fracture Mechanics
of Rock}, ed. B.K. Atkinson (Academic Press, London, 1987), p. 241
\item{2.} H. Van Damme, E. Alsac and C. Laroe, C.R. Acad. Sci.
S\' erie II {\bf 309} (1989) 11; H. Van Damme,
in: {\it The Fractal Approach to Heterogeneous Chemistry},
ed. D. Avnir (Wiley, New York, 1989) p.199; E. Lemaire, P. Levitz,
G. Daccord, and H. Van Damme, Phys. Rev. Lett. {\bf 67} (1991) 2009.
\item{3.} D.A. Kessler, J. Koplik and H. Levine, Adv. Phys. {\bf 37}
(1988) 255
\item{4.} P. Ossadnik, Physica A {\bf 176} (1991) 454
\item{5.} H.J. Herrmann and J. Kert\' esz, Physica A {\bf 178}
(1991) 227
\item{6.} H.J. Herrmann, in: {\it Proceedings of the NATO workshop
``Growth Patterns in Physical Sciences and Biology''} (1992)
\item{7.} E. Louis and F. Guinea, Europhys. Lett. {\bf 3} (1987) 871;
E. Louis, F. Guinea and F. Flores, in: {\it Fractals in Physics},
eds. L. Pietronero and E. Tossatti (Elsevier, Amsterdam, 1986)
\item{8.} P. Meakin, G. Li, L.M. Sander, E. Louis and F. Guinea,
J. Phys. A {\bf 22} (1989) 1393;
E.L. Hinrichsen, A. Hansen and S. Roux, Europhys. Lett.
{\bf 8} (1988) 747
\item{9.} S. Roux and E. Guyon, J. Physique Lett. {\bf 46} (1985) L999
\item{10.} H.J. Herrmann and S. Roux (eds.) {\it Statistical Models
for the Fracture of Disordered Media} (North Holland, Amsterdam, 1990).
\item{11.} The beams are assumed to have the same shape and
the same elastic behavior, governed by three material
dependent constants $a=l/(EA)$, $b=l/(GA)$
and $c=l^3/(EI)$ as described in Ref.~10. Here $l$ is the beam length
(assumed to be unity),
$E$ and $G$ are the Young and shear moduli, $A$ is the area of
the beam section and $I$ is the moment of inertia for flexion.
We used for all simulations the values $a=1.0$, $b=0.017$ and $c=12.0$.
The constant pressure had a value of $0.01$.
\item{12.} For example for a horizontal beam between sites $i$
and $j$ we have for the longitudinal force acting at site $i$,
$-F_i = \alpha (x_i -x_j)$,
for the shear force
$-S_i = \beta (y_i - y_j) +{\beta\over 2}l
(\varphi_i +\varphi_j)$,
and for the flexural torque at site $i$,
$-M_i ={\beta\over 2}l(y_i -y_j +l\varphi_j) +
\delta l^2(\varphi_i-\varphi_j).$
We have used the abbreviations $\alpha = 1/a$,
$\beta=1/(b+c/12)$ and $\delta=\beta(b/c+1/3)$
with respect to the definitions of Ref.~10.
\item{13.} We used in this work
the value $\varepsilon =10^{-10}$ for the relaxation
stopping criterion, see Eq.~(47) in:
G. G. Batrouni and A. Hansen, J. Stat. Phys.
{\bf 52}, 747 (1988).
\item{14.} In this case only a single crack is generated. More
generally, bonds not belonging to the crack surface
can be allowed to break. For these bonds the external pressure does not
act directly since the fluid is supposed to invade only the connected hole.
This leads to further complication of the model which will not be
considered here.
\item{15.} L. Niemeyer, L. Pietronero and H.J. Wiesmann, Phys.
Rev. Lett. {\bf 52} (1984) 1033.
\item{16.} In the case of vanishing cohesion force for tension,
$f_{\rm coh}=0$, there exists always active bonds (i.e. bonds
which can break) with
$p_{ij}>0$ since an infinitesimal external load is sufficient to break
a bond. When $f_{\rm coh}>0$, however, a minimum initial crack length
is required to initiate the fracture process. Since the external load
increases with the number of broken bonds, at which the force dipoles
are applied, the process continues indefinitely.
\item{17.} The calculations were performed on
IBM RS/6000-320H workstations
using for each system of $L = 150$ roughly
100 hours to break $N = 650$ beams.
\item{18.} For a single sample the radius of gyration is usually
defined as
$$R_G^2(N)={1\over N}\sum_{i=1}^N({\bf r}_i - {\bf r}_0)^2, \qquad
 {\bf r}_0 = {1\over N}\sum_{i=1}^N {\bf r}_i,$$
with ${\bf r}_i$ being the position vector of
the $i$th broken beam with respect to the undistorted lattice.
\item{19.} S. Havlin and R. Nossal, J. Phys. {\bf A17 }, L427 (1984).
\item{20.} For an arbitrary broken beam $\ell$ is defined
           as the smallest number of consecutive broken beams
           which one has to
           cross in order to reach the first broken beam
           (center of the lattice).

\item{21.} A. Bunde and S. Havlin (eds.) {\it Fractals and
Disordered Systems} (Springer-Verlag Berlin Heidelberg, 1991),pp.133.
\item{22.} B.R. Lawn and T.R. Wilshaw, {\it Fracture of brittle
           solids} (Cambridge University Press, London 1975).
\item{23.} R.G. Hahn, {\it Bruchmechanik} (Teubner, Stuttgart 1976).
\item{24.} The dominant role of heterogeneities for fracture of
rocks was discussed in recent numerical studies:
M. Sahimi and S. Arbabi, Phys. Rev. Lett. {\bf 68} (1992) 608;
M. Sahimi, M.C. Robertson and C.G. Sammis, Phys.
Rev. Lett. {\bf 70} (1993) 2186.
\vfill\eject
\bigskip
\noindent{\bf Figure Captions}
\medskip
\item{\bf Fig. 1} The beam model on the square lattice. (a)
            A beam connecting sites $i$ and $j$ is shown to
            display the rotational displacements $\varphi$
            at both sites. (b) Diagram of the pair of external
            forces $\vec{F}$ (arrows)
            applied at sites $i$ and $j$ corresponding to
            the initially removed beam.
\medskip
\item{\bf Fig. 2a} A typical crack obtained with a beam model for
            hydraulic fracturing at constant pressure when $\eta=1$,
            on a square lattice of $200\times 200$ sites and
            free external boundary conditions.  The crack
            consists of $310$ broken beams, and only beams
            under tension are allowed to break. The different
            colors represent the intensity of the
            hydrostatic stress field $\vert f_{ij} \vert $.
            The full range for the stress field is
            linearly mapped onto twenty color-circle cycles starting
            at Magenta going through Blue, Cyan, Green, Yellow and
            ending at Red.

\medskip
\item{\bf Fig. 2b} Same as in Fig. 2a for $\eta=0$, on a square lattice
                  of $150\times 150$ sites.
                  This crack consists of $680$ broken beams.
\medskip
\item{\bf Fig. 3} Plot of the number of broken beams $N$
                  as a function of the radius of gyration $R_G$
                  of the cracks obtained when $\eta=0$ and
                  free boundary conditions. Averages
                  over $27$ samples were performed. The inset shows
                  the successive slopes $d_f$ of the data, and indicate
                  an average value $d_f=1.56\pm 0.05$.
\medskip
\item{\bf Fig. 4} Plot of the crack opening volume $V$ as a function
                  of the radius of gyration $R_G$, obtained by averaging
                  over 12 samples using free boundary conditions
                  (triangles) and averaging 4 samples using
                  periodic boundary conditions (squares).
                  For comparison, we also show the results for a straight
                  crack (circles)
                  obtained in the case $\eta\to\infty$ under free
                  boundary conditions.
                  In all three cases
                  we find the behavior $V\sim R_G^2$.

\end